\documentclass[sigconf]{acmart}

\AtBeginDocument{%
  }

\setcopyright{acmlicensed}

\copyrightyear{2026}
\acmYear{2026}
\setcopyright{cc}
\setcctype{by-nc-nd}
\acmConference[ISWC '26]{Proceedings of the 2026 ACM International Symposium on Wearable Computers}{October 11--15, 2026}{Shanghai, China}
\acmBooktitle{Proceedings of the 2026 ACM International Symposium on Wearable Computers (ISWC '26), October 11--15, 2026, Shanghai, China}
\acmDOI{10.1145/3830727.3834841}
\acmISBN{979-8-4007-2872-3/2026/10}

\begin{document}

\title{SkinSpline: A Body-Attached Skeleton-Supported Haptic Interface for Continuous Skin Deformation through Physical Interpolation}

\author{Liwen He}
\orcid{0000-0003-0715-9252}
\authornote{Both authors contributed equally to the paper.}
\affiliation{%
    \department{Academy of Arts \& Design}
    \department{The Future Laboratory}
    \institution{Tsinghua University}
    \city{Beijing}
    \state{}
    \country{China}}
\email{helw24@mails.tsinghua.edu.cn}

\author{Xinyuan Wang}
\orcid{0000-0003-4532-9267}
\authornotemark[1]
\affiliation{%
    \department{Department of Engineering Mechanics, CNMM and AML}
    \institution{Tsinghua University}
    \city{Beijing}
    \state{}
    \country{China}}
\email{pch172@qq.com}

\author{Jiachen Du}
\orcid{0000-0001-6444-9112}
\affiliation{%
    \department{The Future Laboratory}
    \institution{Tsinghua University}
    \city{Beijing}
    \state{}
    \country{China}}
\email{dujiachen1226@163.com}

\author{Zixin Chen}
\orcid{0009-0000-7426-9174}
\affiliation{%
    \department{School of Mechanical Engineering and Automation}
    \institution{Beihang University}
    \city{Beijing}
    \state{}
    \country{China}}
\email{zixinchen@buaa.edu.cn}

\author{Yun Wang}
\authornote{Corresponding author.}
\orcid{0000-0001-9847-2636}
\affiliation{%
    \department{School of New Media Art and Design}
    \department{State Key Lab of Virtual Reality Technology and Systems}
    \institution{Beihang University}
    \city{Beijing}
    \state{}
    \country{China}
}
\email{wang_yun@buaa.edu.cn}

\renewcommand{\shortauthors}{He et al.}
\renewcommand{\shorttitle}{SkinSpline}

\begin{abstract}

We present SkinSpline, a body-attached skeleton-supported haptic interface that renders continuous skin deformation through physical interpolation of sparse mechanical actuation. SkinSpline combines a low-resolution array of rack-and-pinion linear actuators with an elastic interlocking skeleton that transforms discrete actuator motions into smooth surface deformation, enabling continuous cutaneous feedback without dense actuator arrays. 
The system includes a modular hardware architecture, a configurable control pipeline, and a visual interface supporting real-time configuration and actuation. We demonstrate SkinSpline through multiple scenarios, including wave rendering, video-synchronized rhythmic touch, visually driven water-wave feedback in VR, and sensor-based remote touch reproduction. SkinSpline explores an alternative approach to continuous on-body haptic rendering by leveraging structural coupling between sparse actuation and deformable surfaces.

\end{abstract}



\begin{teaserfigure}
  \includegraphics[width=\textwidth]{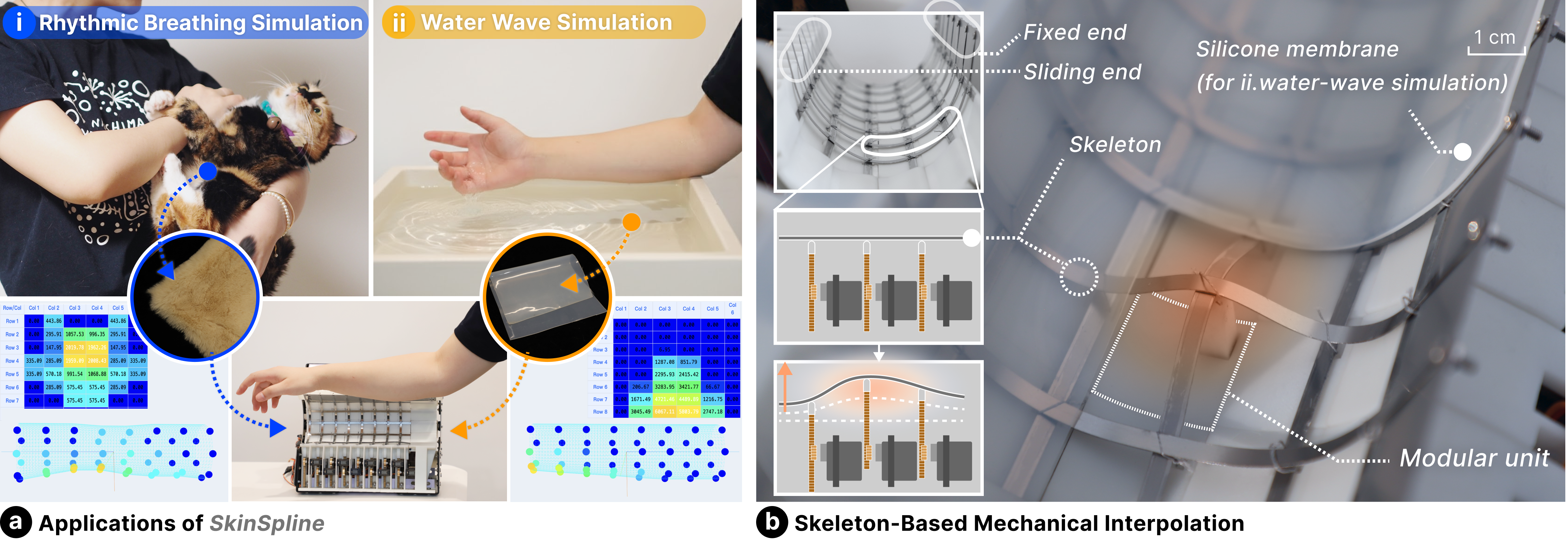}
  \caption{
  SkinSpline. (a) Rhythmic breathing and water-wave demonstrations. (b) Independently controlled units deform an interlocking elastic skeleton, physically interpolating sparse actuation into continuous surface deformation. The silicone membrane is used for water-wave feedback.}
  \Description{Overview of SkinSpline. Part (a) shows rhythmic breathing and water-wave demonstrations, their spatial input matrices, and the forearm-scale prototype. Part (b) shows the interlocking steel skeleton, modular linear actuation units, fixed and sliding strip ends, and the silicone covering used for water-wave feedback. Diagrams illustrate how actuator displacement produces a continuous curved surface.}
  \label{fig:teaser}
\end{teaserfigure}


\maketitle

\section{Introduction}

Body-attached haptic interfaces can support immersive, affective, and assistive interaction by delivering tactile feedback directly on the skin~\cite{obrist2016sensing,chien2021flowing}. While many on-body systems rely on vibration or localized point stimulation, a range of tactile experiences, such as flowing water, rhythmic pressure, or distributed surface motion, require spatially continuous skin deformation across larger body areas~\cite{goetz2023dynamic,sung2024hapticpilot}. Rendering such continuous cutaneous feedback remains challenging for wearable devices.



Existing approaches face a trade-off between spatial continuity and controllability. Array-based wearable haptic interfaces, including vibrotactile, pin-based, electroadhesive, dielectric-elastomer, and microfluidic displays, provide spatially distributed stimulation but often rely on discrete stimulation points~\cite{yu2019skin,shen2023fluid,tan2025toward,youn2025skin,he2025understanding}. Perceptual interpolation and phantom sensations can support motion cues~\cite{massie1994phantom,kato2010basic,fang2025hapemoji}, but sparse arrays may still feel fragmented when rendering large-area pressure or sustained surface deformation. In contrast, soft pneumatic or fluidic interfaces can produce smoother contact over continuous materials~\cite{shao2020surfaceflow,nunez2022large}, but they often provide limited local controllability, dynamic range, or scalability for complex spatial patterns~\cite{yu2025soft,grasso2023fully,israr2011tactile}.

We present \textbf{SkinSpline}, a body-attached skeleton-supported haptic interface for rendering continuous skin deformation through physical interpolation. SkinSpline combines a low-resolution array of rack-and-pinion linear actuators with an interlocking elastic skeleton. Instead of increasing actuator density, the skeleton mechanically couples neighboring actuation points and transforms discrete vertical displacements into smooth, spline-like surface deformation. This mechanism supports continuous cutaneous feedback while retaining addressable actuation across a forearm-scale surface.

SkinSpline integrates a modular hardware architecture, a calibration-enabled control pipeline, and a visual configuration interface for designing and monitoring spatial haptic patterns. We demonstrate the system through four input modes: algorithmic wave rendering, video-synchronized rhythmic touch, VR-based water-wave feedback, and sensor-driven remote touch reproduction. Through these demonstrations, SkinSpline explores how physical interpolation can provide an alternative path for continuous on-body haptic rendering with sparse mechanical actuation.

\section{Related Work and Positioning}

\textbf{Discrete arrays and perceptual interpolation.} Spatial haptic interfaces use vibrotactile, pin-based, electroadhesive, dielectric-elastomer, pneumatic, or microfluidic actuator arrays~\cite{yu2019skin,shen2023fluid,tan2025toward,youn2025skin,he2025understanding}. Sparse layouts can convey apparent motion through phantom sensations or sequential activation~\cite{massie1994phantom,kato2010basic,fang2025hapemoji}, but still output at separate contact points. Shape-Kit likewise produces expressive on-body haptics at discrete regions~\cite{shapekit2025}.

\textbf{Continuous surfaces and physical interpolation.} Pneumatic, fluidic, and deformable-surface interfaces provide compliant, continuous contact, although large-area systems often limit independent spatial control~\cite{shao2020surfaceflow,nunez2022large,grasso2023fully}. Embedding discrete actuators in elastic membranes or fluid-filled structures instead uses mechanical coupling to interpolate between actuation points~\cite{phung2020haptic,grasso2023fully}. SkinSpline applies this principle through an interlocking elastic skeleton that converts sparse linear displacements into continuous two-dimensional deformation.

\section{System Design: SkinSpline}
\begin{figure}
    \centering
    \includegraphics[width=1\linewidth]{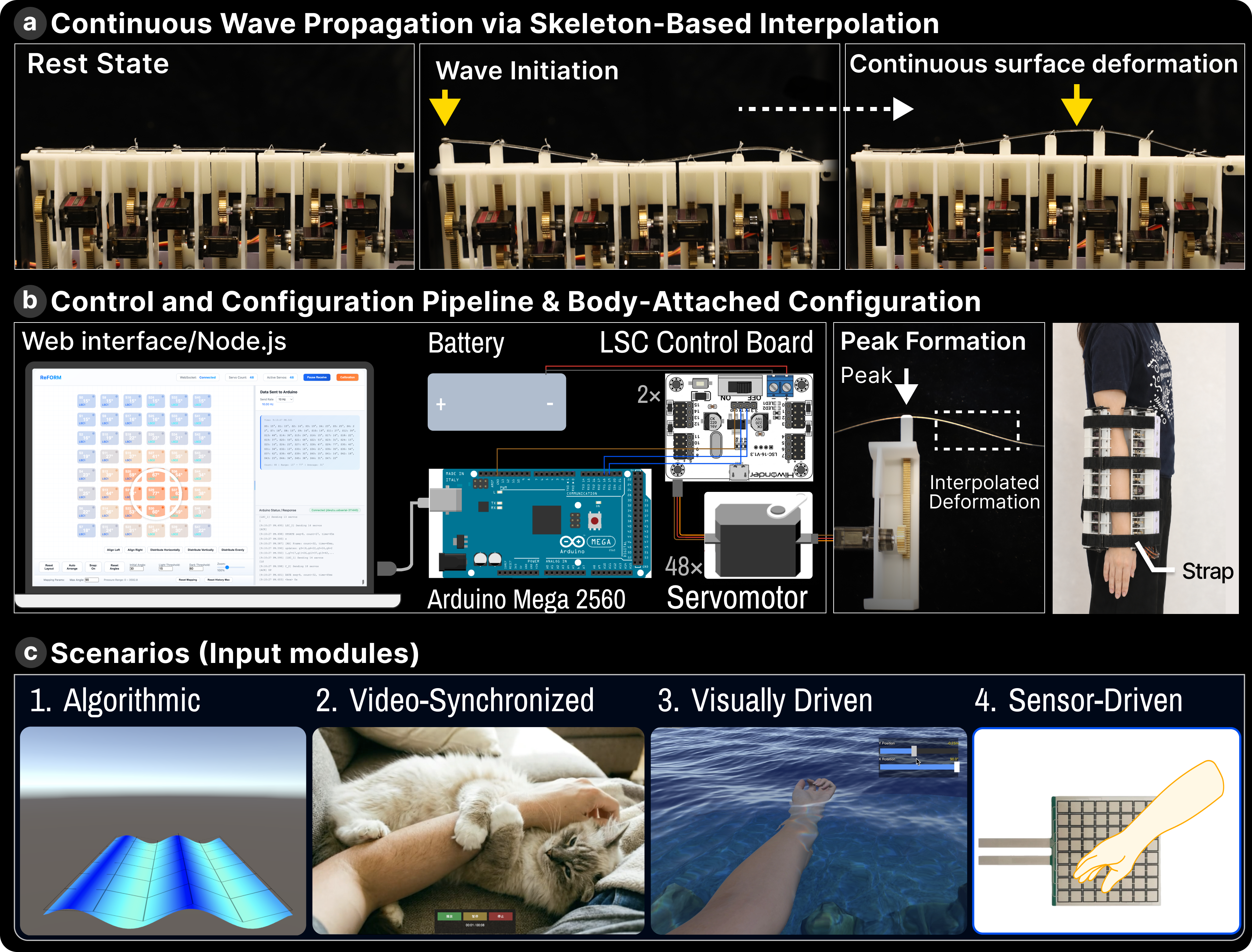}
    \caption{SkinSpline overview: (a) continuous wave rendering, (b) control pipeline and body-attached configuration, and (c) four input modules.}
    \label{fig:overview}
\end{figure}

SkinSpline implements physical interpolation in a forearm-scale, body-attached prototype. The system comprises 48 independently controlled rack-and-pinion linear actuation units arranged in a $6 \times 8$ grid beneath an interlocking elastic skeleton, forming a continuous semi-cylindrical contact surface. Interchangeable coverings modify the contact properties, while a layered control system maps spatial input data to per-actuator displacement commands (Figure~\ref{fig:overview}).

\subsection{Physical Interface Design}

\subsubsection{Actuation Unit Design}
Each actuation unit uses an MG90S micro-servo and a brass rack-and-pinion transmission to produce up to 33\,mm of vertical displacement. The servo operates at 6\,V, weighs 12\,g, provides a rated torque of 2.8\,kg$\cdot$cm, and has a nominal speed of 0.09\,s per 60$^\circ$. Its layered assembly allows an individual unit to be serviced or replaced without rebuilding the full array (Figure~\ref{fig:overview}b).

\subsubsection{Deformable Surface Interface}
The active surface is a semi-cylinder ($R=45$\,mm, $H=200$\,mm) containing a $6 \times 8$ grid of actuation points. A user may place the forearm within the interface for supported use or secure the frame to the arm with straps for body-attached operation. The current prototype measures approximately $140 \times 255 \times 274$\,mm and has an estimated component mass of 2.8\,kg excluding the external power supply; the 48 servos account for 576\,g, with the remaining mass primarily arising from the 3D-printed frame and transmission components.

An orthogonal, interlocking skeleton fabricated from SUS301 stainless steel connects the actuator tips. Flexible circumferential strips (0.15\,mm thick) conform around the forearm, while stiffer longitudinal strips (0.2\,mm thick) maintain continuity along the surface. Flexible linkages transmit vertical displacement while permitting local rotation and in-plane movement. The longitudinal strips terminate in sliding guides with spring tensioning, accommodating arc-length changes and reducing buckling during deformation. Silicone, spandex, or plush coverings can be placed over the skeleton to vary compliance, friction, and texture (Figure~\ref{fig:teaser}).

\subsubsection{Skeleton-Based Mechanical Interpolation}
Between adjacent actuation points, an unloaded elastic strip follows the Euler-Bernoulli relation
\begin{equation}
EI\frac{d^4y}{dx^4}=0,
\end{equation}
where $E$ is the Young's modulus, $I$ is the second moment of area, and $y(x)$ is the beam deflection. The resulting cubic beam segments maintain continuity of displacement, slope, and bending moment across adjacent nodes. The orthogonal skeleton therefore acts as a two-dimensional physical spline, transforming the independently specified actuator heights into a continuous surface without numerical interpolation between nodes.

\subsection{Control System}

SkinSpline features a configurable control system managing data flow, actuator coordination, and real-time execution.

\subsubsection{System Architecture and Data Flow}
The control pipeline comprises an upstream application, a Node.js server, an Arduino Mega 2560, and two LSC servo controller boards. Upstream modules transmit a frame of 48 target angles to the server through WebSocket. The server bounds the commands and packs changed values into a compact binary frame sent to the Arduino over USB serial at 115200\,bps. Each actuator has a unique global ID (0-47): the Arduino routes IDs 0-23 to the first LSC board and IDs 24--47 to the second, with each board controlling 24 servos over a dedicated 9600-bps serial channel. Thus, all 48 units are independently addressable rather than operated as fixed groups.

The current implementation is configured for a 100-ms command interval (10\,Hz) and a nominal servo movement time of 120\,ms. To reduce traffic on the LSC links, the server transmits channels whose targets change by at least 2$^\circ$ and forces a complete 48-channel refresh every second. This configuration has been used with the present array; larger arrays would require additional controller hardware or distributed embedded gateways.

\subsubsection{Mapping, Calibration, and Interface}
Each upstream module maps its spatial data to the configured actuator layout before transmission. For example, the wave-rendering module linearly maps sampled surface heights to servo angles and limits the commanded range to 60--210$^\circ$; the server and embedded controller additionally clamp decoded commands to the servo's nominal 0--270$^\circ$ range. A saved initial-position offset is then applied to each actuator, compensating for assembly variation and adapting the resting surface to different forearm geometries or scene-specific contact conditions.

The web-based interface supports layout configuration, per-actuator calibration, and real-time monitoring. Users can drag individual actuator nodes on a 2D canvas to reproduce the physical arrangement and assign each node to its hardware channel. In calibration mode, each actuator can be tested and adjusted independently to establish an initial position that fits the user's arm or the intended surface geometry. Layouts, channel assignments, calibration offsets, and motion parameters can be saved as profiles and reloaded for subsequent users or scenarios. During operation, the same canvas visualizes the commanded $6 \times 8$ height and angle matrices and reports the connection state.

The parametric waveform generator is an upstream software module connected through the same control interface and WebSocket pipeline. Video-, VR-, and sensor-driven applications replace this input module while retaining the configured layout, calibration profile, server, addressing, and actuation layers. Incomplete binary frames are discarded after a 50-ms parser timeout, and a reset command returns all actuators to 0$^\circ$. The current actuation is open-loop and does not detect contact force or motor stall; accordingly, the prototype is operated under researcher supervision. Actuation power is supplied independently of the host computer by two 3S1P LiFePO$_4$ battery packs (9.6\,V nominal, 1500\,mAh each; 28.8\,Wh combined). The batteries connect to the power inputs of the two LSC boards, which distribute power through their servo rails to the 48 actuators. The computer provides USB serial communication to the Arduino Mega 2560 but does not supply the LSC boards or servo array.

\subsection{Demonstration Scenarios}

The four demonstrations use different upstream input sources while sharing the same downstream control and rendering pipeline (Figure~\ref{fig:overview}c).

\textit{Demo 1: Algorithmic Waveform Demonstration}
A parametric generator produces sinusoidal, composite, and pulsed patterns. Amplitude, wavelength, propagation direction, and phase determine the 48 target values, producing spatially propagating deformation across the surface.

\textit{Demo 2: Video-Synchronized Rhythmic Touch Simulation}
A Unity application synchronizes video playback with a prerecorded $8 \times 6$ pressure sequence sampled at 20\,Hz. The sequence is mapped to SkinSpline to reproduce periodic forearm deformation aligned with the observed rhythmic contact.

\textit{Demo 3: Visually Driven Water Wave Simulation}
A Unity-based VR application samples water depth at multiple points on a forearm proxy and converts the values into a $6 \times 8$ input matrix. SkinSpline renders deformation patterns that vary with virtual immersion, orientation, and simulated wave motion.

\textit{Demo 4: Sensor-Driven Remote Touch Reproduction}
An $8 \times 8$ thin-film pressure sensor array captures spatial touch, from which a $6 \times 8$ region is mapped to SkinSpline. The server visualizes the incoming matrix and forwards it through the shared pipeline for streamed spatial touch reproduction.

\section{Discussion and Limitations}

SkinSpline demonstrates a structural approach to continuous cutaneous rendering: independently controlled, sparse actuator displacements are mechanically interpolated by an interlocking elastic skeleton into a continuous forearm-scale surface. The shared control pipeline enables the same physical interface to render traveling deformation, rhythmic pressure, simulated environmental effects, and remotely captured touch. Beyond the demonstrated forearm configuration, this mechanism may inform future body-contact or embedded interfaces in which spatially continuous deformation is preferable to isolated point stimulation (Appendix \ref{app:conceptual-extensions}).

This work focuses on system design and application demonstrations rather than quantitative or perceptual evaluation. The current prototype is rigid, has an estimated mass of 2.8\,kg excluding its batteries, and has only been used for short-term, supervised interaction. Although straps allow it to remain attached during arm movement, its mobility and extended-wear comfort have not been evaluated. The 48-actuator system operates with command-level position control and does not measure contact force, actual surface displacement, or motor stall. Its performance beyond the present forearm-scale configuration has also not been validated. Future work will characterize displacement accuracy, surface continuity, latency, force output, power consumption, and failure behavior, together with user studies of perceived continuity, comfort, and body fit. Reducing structural and actuator mass will be important for more mobile and extended use.

\begin{acks}
This work was supported by the National Natural Science Foundation of China (NSFC) under Grant No. 62403030.
\end{acks}

\bibliographystyle{ACM-Reference-Format}
\bibliography{0-reference}

\appendix

\section{Conceptual Extensions of the SkinSpline Mechanism}
\label{app:conceptual-extensions}

\begin{figure}[t]
    \centering
    \includegraphics[width=\linewidth]{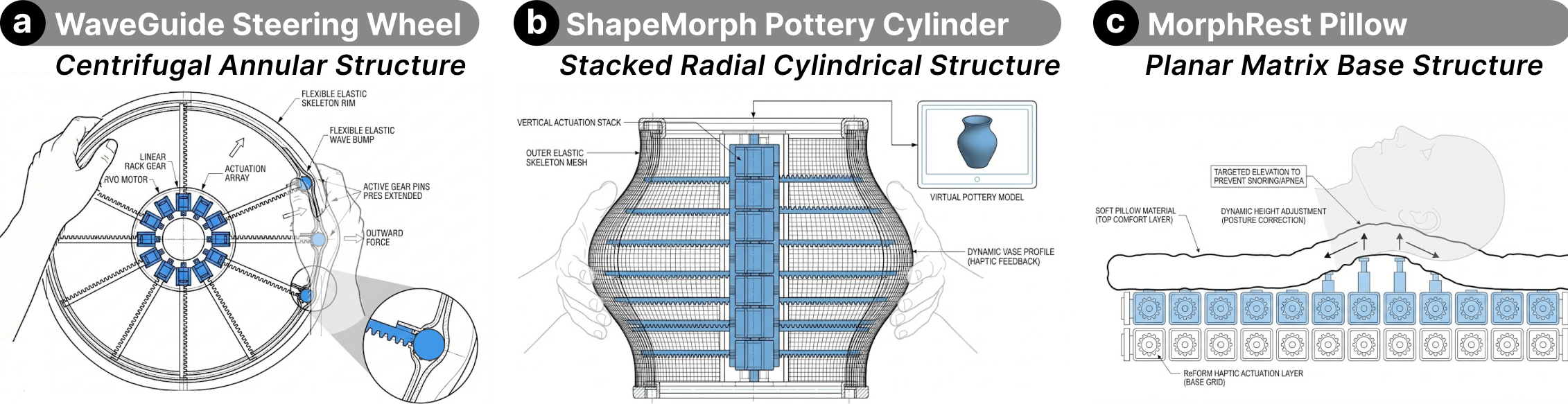}
    \caption{Conceptual extensions of SkinSpline's skeleton-supported deformation mechanism to steering, handheld, and body-support interfaces. These examples illustrate potential form factors rather than implemented or evaluated applications.}
    \Description{Concept illustrations applying the SkinSpline deformation mechanism to steering, handheld, and body-support interfaces.}
    \label{fig:conceptual-extensions}
\end{figure}

Figure~\ref{fig:conceptual-extensions} illustrates three conceptual extensions of SkinSpline's skeleton-supported deformation mechanism. By adapting the geometry, boundary conditions, and contact configuration of the interlocking skeleton, the same physical-interpolation principle could potentially support steering, handheld, and body-support interfaces. These concepts are included to indicate possible directions for future exploration; they have not been implemented or evaluated in the present work.

\end{document}